\documentclass{article}
\usepackage{amsfonts}

\usepackage{graphicx}
\usepackage{amsmath}

\input{tcilatex}

\begin{document}

\title{New spherically symmetric monopole and regular solutions in
Einstein-Born-Infeld theories}
\author{D.J. Cirilo Lombardo \\
Bogoliubov Laboratory of Theoretical Physics \\
Joint Institute of Nuclear Research, \\
Dubna, Moscow Region, 141980, Russia\thanks{%
e-mails:diego@thsun1.jinr.ru diego77jcl@yahoo.com.}}
\maketitle

\begin{abstract}
In this work a new asymptotically flat solution of the coupled
Einstein-Born-Infeld equations for a static spherically symmetric space-time
is obtained. When the intrinsic mass is zero the resulting spacetime is
regular everywhere, in the sense given by B. Hoffmann and L. Infeld in 1937,
and the Einstein-Born-Infeld theory leads to the identification of the
gravitational with the electromagnetic mass. This means that the metric, the
electromagnetic field and their derivatives have not discontinuities in all
the manifold. In particular, there are not conical singularities at the
origin, in contrast to well known monopole solution studied by B. Hoffmann
in 1935. The lack of uniqueness of the action function in
Non-Linear-Electrodynamics is discussed.
\end{abstract}

\section{\protect\bigskip Introduction and results:}

\noindent

The four dimensional solutions with spherical symmetry of the Einstein
equations coupled to Born-Infeld fields have been well studied in the
literature$^{1-4}$. In particular, the electromagnetic field of the Born
Infeld monopole, in contrast to Maxwell counterpart, contributes to the ADM\
mass of the system (it is, the four momentum of asymptotic flat manifolds).
B. Hoffmann was the first who studied such static solutions in the context
of the general relativity with the idea of to obtain a consistent
particle-like model$^{2}$ . Unfortunately, these static Einstein-Born-Infeld
models generate conical singularities at the origin$^{2-3}$ that cannot be
removed as in global monopoles or other non-localized defects of the
spacetime$^{5-6}$. With the existence of this type of singularities in the
space-time of the monopole we can not identificate the gravitational with
the electromagnetic mass. In this work a \textit{new} static spherically
symmetric solution with Born-Infeld charge is obtained. The new metric, when
the intrinsic mass of the system is zero, is \textit{regular} everywhere in
the sense that was given by B. Hoffmann and L. Infeld$^{3}$ in 1937 and the
EBI\ theory leads to identification of the gravitational with the
electromagnetic mass. This means that the metric, the electromagnetic field
and their derivatives have not singularities and discontinuities in all the
manifold. The fundamental feature of this solution is the lack of conical
singularities at the origin. A distant observer will associate with this
solution an electromagnetic mass that is a twice of the mass of the
electromagnetic geon founded by M. Demianski$^{4}$ in 1986 .\ The
energy-momentum tensor and the electric field are both regular with zero
value at the origin and new parameters appear, given to the new metric
surprising behaviours. The used convention$^{7-8}$ is the \textit{spatial }%
of Landau and Lifshitz (1962), with signatures of the metric, Riemann and
Einstein tensors all positives (+++) .

The plan of this paper is as follows: in Section 2 we give a short
introduction to the Born-Infeld theory: propierties and principal features.
In Section 3 the regularity condition as was given by B. Hoffmann and L.
Infeld$^{3}$ in 1937 . Sections 4, 5, 6 and 7 are devoted to found the new
solution and to analyze its propierties. Finally, the conclusion and
comments of the results are presented in section 8.

\section{\protect\bigskip The Born-Infeld theory:}

The most significative non-linear theory of electrodynamics is, by
excellence, the Born-Infeld theory$^{1,9}$. Among its many special
properties is an exact SO(2) electric-magnetic duality invariance. The
Lagrangian density describing Born-Infeld theory (in arbitrary spacetime
dimensions) is
\begin{equation}
\mathcal{L}_{BI}=\sqrt{-g}L_{BI}=\frac{b^{2}}{4\pi }\left\{ \sqrt{-g}-\sqrt{%
\left| \det (g_{\mu \nu }+b^{-1}F_{\mu \nu })\right| }\right\}
\end{equation}
where $b$ is a fundamental parameter of the theory with field dimensions. In
open superstring theory$^{10}$, for example, loop calculations lead to this
Lagrangian with $b^{-1}=2\pi \alpha ^{\prime }$ ($\alpha ^{\prime }\equiv \ $%
inverse of the string tension)\ . In four spacetime dimensions the
determinant in (1) may be expanded out to give
\begin{equation}
L_{BI}=\frac{b^{2}}{4\pi }\left\{ 1-\sqrt{1+\frac{1}{2}b^{-2}F_{\mu \nu
}F^{\mu \nu }-\frac{1}{16}b^{-4}\left( F_{\mu \nu }\widetilde{F}^{\mu \nu
}\right) ^{2}}\right\}
\end{equation}
which coincides with the usual Maxwell Lagrangian in the weak field limit.

It is useful to define the second rank tensor $P^{\mu \nu }$ by
\begin{equation}
P^{\mu \nu }=-\frac{1}{2}\frac{\partial L_{BI}}{\partial F_{\mu \nu }}=\frac{%
F^{\mu \nu }-\frac{1}{4}b^{-2}\left( F_{\rho \sigma }\widetilde{F}^{\rho
\sigma }\right) \,\widetilde{F}^{\mu \nu }}{\sqrt{1+\frac{1}{2}b^{-2}F_{\rho
\sigma }F^{\rho \sigma }-\frac{1}{16}b^{-4}\left( F_{\rho \sigma }\widetilde{%
F}^{\rho \sigma }\right) ^{2}}}
\end{equation}
(so that $P^{\mu \nu }\approx F^{\mu \nu }$ for weak fields) satisfying the
electromagnetic equations of motion
\begin{equation}
\nabla _{\mu }P^{\mu \nu }=0
\end{equation}
which are highly non linear in $F_{\mu \nu }$. The energy-momentum tensor
may be written as
\begin{equation}
T_{\mu \nu }=\frac{1}{4\pi }\left\{ \frac{F_{\mu }^{\,\ \ \lambda }F_{\nu
\lambda }+b^{2}\left[ \mathbb{R}-1-\frac{1}{2}b^{-2}F_{\rho \sigma }F^{\rho
\sigma }\right] g_{\mu \nu }}{\mathbb{R}}\right\}
\end{equation}
\begin{equation*}
\mathbb{R}\equiv \sqrt{1+\frac{1}{2}b^{-2}F_{\rho \sigma }F^{\rho \sigma }-%
\frac{1}{16}b^{-4}\left( F_{\rho \sigma }\widetilde{F}^{\rho \sigma }\right)
^{2}}
\end{equation*}
Although it is by no means obvious, it may verified that equations (3)-(5)
are invariant under electric-magnetic rotations of duality $%
F\longleftrightarrow \ast G$. We can show that the SO$\left( 2\right) $
structure of the Born-Infeld theory is more easily seen in quaternionic form$%
^{11-12}$
\begin{equation*}
\frac{1}{R}\left( \sigma _{0}+i\sigma _{2}\overline{\mathbb{P}}\right) L=%
\mathbb{L}
\end{equation*}
\begin{equation*}
\frac{\mathbb{R}}{\left( 1+\overline{\mathbb{P}}^{2}\right) }\left( \sigma
_{0}-i\sigma _{2}\overline{\mathbb{P}}\right) \mathbb{L}=L
\end{equation*}
\begin{equation*}
\overline{\mathbb{P}}\equiv \frac{\mathbb{P}}{b}
\end{equation*}
where we defined
\begin{equation*}
L=F-i\sigma _{2}\widetilde{F}
\end{equation*}
\begin{equation*}
\mathbb{L}=P-i\sigma _{2}\widetilde{P}
\end{equation*}
the pseudoescalar of the electromagnetic tensor $F^{\mu \nu }$%
\begin{equation*}
\mathbb{P}=-\frac{1}{4}F_{\mu \nu }\widetilde{F}^{\mu \nu }
\end{equation*}
and $\sigma _{0}\,$, $\sigma _{2}$ the well know Pauli matrix.

In flat space, and for purely electric configurations, the Lagrangian (2)
reduces to
\begin{equation*}
L_{BI}=\frac{4\pi }{b^{2}}\left\{ 1-\sqrt{1-b^{-2}\overrightarrow{E^{2}}}%
\right\}
\end{equation*}
so there is an upper bound on the electric field strength $\overrightarrow{E}
$
\begin{equation}
\left| \overrightarrow{E}\right| \leq b
\end{equation}

\section{\protect\bigskip The regularity condition}

The new field theory initiated in 1934 by M. Born$^{9}$ introduces in the
classical equations of the electromagnetic field a characteristic length $%
r_{0}$ representing the radius of the elementary particle through the
relation
\begin{equation*}
r_{0}=\sqrt{\frac{e}{b}}
\end{equation*}
where $e$ is the elementary charge and $b$ the fundamental field strength
entering in a non-linear Lagrangian function. It was originally thought that
the Lagrangian (1) was the simplest choice which would lead to a finite
energy for an electric particle. This is, however, not the case. It is
possible to find an infinite number of quite different action functions,
each giving simple algebraic relations between the fields and each leading
to a finite energy for an electric particle.

In 1937 B. Hoffmann and L. Infeld$^{3}$ introduce a regularity condition on
the new field theory of M. Born$^{9}$with the main idea of to solve the lack
of uniqueness of the function action. They have already seen that the
condition of regularity of the field gives the restriction in the
spherically symmetric electrostatic case $E_{r}=0$ for $r=0$.

In the general theory they applyed the regularity condition not only to the $%
F_{\mu \nu }$ field but also to the $g_{\mu \nu }$ field. The regularity
condition for the general theory was that:

\textit{Only those solutions of the fields equations may have physical
meaning for which space-time is everywhere regular and for which the }$%
F_{\mu \nu }$\textit{\ and the }$g_{\mu \nu }$\textit{\ fields and those of
their derivatives which enter in the field equations and the conservation
laws exist everywhere.}

In the general theory of the relativity the spherically symmetric solution
of the purely gravitational field equations is given by the Schwarzschild
line element
\begin{equation*}
ds^{2}=-Adt^{2}+A^{-1}dr^{2}+r^{2}\left( d\theta ^{2}+\sin ^{2}\theta
\,d\varphi ^{2}\right)
\end{equation*}
\begin{equation*}
A\equiv 1-\frac{2M}{r}
\end{equation*}
where $(-2M)$ is a constant of integration $\ M$ have having the
significance of the gravitational mass of the body source of the field (we
take the gravitational constant $G=1$). This line element has an essential
singularity at $r=0$ and does satisfy the regularity condition.

In the general relativity form of the original new field theory the
requeriment that there be no infinities in the $g_{\mu \nu }$ forces the
identification of gravitational with electromagnetic mass. In$^{3}$ B.
Hoffmann and L. Infeld have used for such identification the line element of
the well known monopole solution studied by B. Hoffmann$^{2}$ in 1935
\begin{equation*}
A\equiv 1-\frac{8\pi }{r}\int_{0}^{r}\left[ (r^{4}+1)^{1/2}-r^{2}\right] dr
\end{equation*}
that is originated by an Einstein-Born-Infeld action as in equation (1).
This line element approximates the Schwarzschild form for $r$ greater than
the electronic radius but avoid the infinities of that line element for $r=0$%
. However is still a singularity of conical type at the pole. When $%
r\rightarrow 0$ the above expression for \ $A$, gives
\begin{equation*}
A\rightarrow (1-8\pi )\equiv \beta
\end{equation*}
so $ds^{2}$ becomes
\begin{equation*}
ds^{2}=-\beta dt^{2}+\beta ^{-1}dr^{2}+r^{2}\left( d\theta ^{2}+\sin
^{2}\theta \,d\varphi ^{2}\right)
\end{equation*}
Thus the radio of the circumference to the radius of a small circle having
its centre at the pole is, in the limit, $2\pi \beta $ and not $2\pi $.
Therefore the origin (it is, at $r=0$) is a conical point and not regular.
Note that, because the conical point, no coordinate can be introduced which
will be non singular at $r=0$ and derivatives are actually undefined at this
point.

This problem with the conical singularities at $r=0$ , that destroy the
regularity condition, makes that in the reference$^{3}$ B. Hoffmann and L.
Infeld change the action of the Born-Infeld form as in equation (1) for
other non-linear Lagrangian of logarithmic type .The new logarithmic action
does not presented such difficulties at $r=0$, and makes that time ago many
people changes the very nice form of the Einstein-Born-Infeld action (1) for
others non-linear Lagrangians that solved the problem of the self-energy of
the electron and the regularity condition given above.

In this work we presented a new spherically symmetric solution of the
Einstein-Born-Infeld equations. The metric, when the intrinsic mass of the
system is zero, is \textit{regular} everywhere in the sense that was given
by B. Hoffmann and L. Infeld$^{3}$ in 1937, and the EBI\ theory leads to
identification of the gravitational with the electromagnetic mass. In this
manner we also show that more strong conditions are needed for to solve the
problem of the lack of uniqueness of the function action.

\section{Statement of the problem:}

We propose the following line element for the static Born-Infeld monopole

\begin{equation}
ds^{2}=-e^{2\Lambda }dt^{2}+e^{2\Phi }dr^{2}+e^{2F\left( r\right) }d\theta
^{2}+e^{2G\left( r\right) }\sin ^{2}\theta \,d\varphi ^{2}
\end{equation}
where the components of the metric tensor are
\begin{equation}
\begin{array}{cccc}
g_{tt}=-e^{2\Lambda } &  &  & g^{tt}=-e^{-2\Lambda } \\
g_{rr}=e^{2\Phi } &  &  & g^{rr}=e^{-2\Phi } \\
g_{\theta \theta }=e^{2F} &  &  & g^{\theta \theta }=e^{-2F} \\
g_{\varphi \varphi }=\sin ^{2}\theta \,e^{2G} &  &  & g^{\varphi \varphi }=%
\frac{e^{-2G}}{\sin ^{2}\theta }
\end{array}
\end{equation}
For the obtention of \ the Einstein-Born-Infeld equations system we use the
Cartan's structure equations method$^{13}$, that is most powerful and direct
where we work with differential forms and in a orthonormal frame (tetrad).
The line element (7) in the 1-forms basis takes the following form
\begin{equation}
ds^{2}=-\left( \omega ^{0}\right) ^{2}+\left( \omega ^{1}\right) ^{2}+\left(
\omega ^{2}\right) ^{2}+\left( \omega ^{3}\right) ^{2}
\end{equation}
were the forms are
\begin{equation}
\begin{array}{cccc}
\omega ^{0}=e^{\Lambda }dt &  & \Rightarrow & dt=e^{-\Lambda }\omega ^{0} \\
\omega ^{1}=e^{\Phi }dr &  & \Rightarrow & dr=e^{-\Phi }\omega ^{1} \\
\omega ^{2}=e^{F\left( r\right) }d\theta &  & \Rightarrow & d\theta
=e^{-F\left( r\right) }\omega ^{2} \\
\omega ^{3}=e^{G\left( r\right) }\sin \theta \,d\varphi &  & \Rightarrow &
d\varphi =e^{-G\left( r\right) }\left( \sin \theta \right) ^{-1}\omega ^{3}
\end{array}
\end{equation}
Now, we follow the standard procedure of the structure equations (Appendix):
\begin{equation}
d\omega ^{\alpha }=-\omega ^{\alpha }\,_{\beta }\wedge \omega ^{\beta }
\end{equation}
\begin{equation}
\mathcal{R}^{\alpha }\,_{\beta }=d\,\omega ^{\alpha }\,_{\beta }+\omega
^{\alpha }\,_{\lambda }\wedge \omega ^{\lambda }\,_{\beta }
\end{equation}
The components of the Riemann tensor are easily obtained from the well know
geometrical relation of Cartan:
\begin{equation*}
\mathcal{R}^{\alpha }\,_{\beta }=R^{\alpha }\,_{\beta \rho \sigma
}\,\,\omega ^{\rho }\wedge \omega ^{\sigma }
\end{equation*}
where we obtain explicitly

\bigskip

\begin{equation}
R^{0}\,_{110}=e^{-2\Phi }\left( \partial _{r}\partial _{r}\Lambda -\partial
_{r}\Phi \,\partial _{r}\Lambda +\left( \partial _{r}\Lambda \right)
^{2}\right)
\end{equation}

\begin{equation*}
R^{2}\,_{112}=e^{-2\Phi }\left( \partial _{r}\partial _{r}F-\partial
_{r}\Phi \,\partial _{r}F+\left( \partial _{r}F\right) ^{2}\right)
\end{equation*}

\begin{equation*}
R^{3}\,_{113}=e^{-2\Phi }\left( \partial _{r}\partial _{r}G-\partial
_{r}\Phi \,\partial _{r}G+\left( \partial _{r}G\right) ^{2}\right)
\end{equation*}

\begin{equation*}
R^{3}\,_{213}\,\,=e^{-\left( F+\Phi \right) }\,\partial _{r}\left(
G-F\right) \,\frac{\cos \theta }{\sin \theta }
\end{equation*}

\begin{equation*}
R^{3}\,_{123}\,\,=e^{-\left( F+\Phi \right) }\,\partial _{r}\left(
G-F\right) \,\frac{\cos \theta }{\sin \theta }
\end{equation*}
\begin{equation*}
R^{3}\,_{223}\,\,=e^{-2\Phi }\partial _{r}G\,\,\partial _{r}F-e^{-2F}
\end{equation*}
\begin{equation*}
R^{0}\,_{330}=e^{-2\Phi }\,\partial _{r}\Lambda \,\,\partial _{r}G
\end{equation*}
\begin{equation*}
R^{0}\,_{220}=e^{-2\Phi }\,\partial _{r}\Lambda \,\,\partial _{r}F
\end{equation*}
From the components of the Riemann tensor we can construct the Einstein
equations
\begin{equation}
G^{0}\,_{0}=-\left( R^{12}\,_{12}+R^{23}\,_{23}+R^{31}\,_{31}\right)
\end{equation}
\begin{equation}
G^{1}\,_{1}=-\left( R^{02}\,_{02}+R^{03}\,_{03}+R^{23}\,_{23}\right)
\end{equation}
\begin{equation}
G^{0}\,_{1}=R^{02}\,_{12}+R^{03}\,_{13}
\end{equation}
\begin{equation}
G^{1}\,_{2}=R^{10}\,_{20}+R^{13}\,_{23}
\end{equation}
explicitly
\begin{equation}
G^{1}\,_{2}=-e^{-\left( F+G\right) }\frac{\cos \theta }{\sin \theta }%
\partial _{r}\left( G-F\right)
\end{equation}

\bigskip

\begin{equation}
G^{0}\,_{0}=e^{-2\Phi }\Psi -e^{-2F}
\end{equation}
where
\begin{equation*}
\Psi =\left[ \partial _{r}\partial _{r}\left( F+G\right) -\partial _{r}\Phi
\,\partial _{r}\left( F+G\right) +\left( \partial _{r}F\right) ^{2}+\left(
\partial _{r}G\right) ^{2}+\partial _{r}F\,\partial _{r}G\right]
\end{equation*}

\bigskip

\begin{equation}
G^{1}\,_{1}=e^{-2\Phi }\left[ \partial _{r}\Lambda \,\partial _{r}\left(
F+G\right) +\partial _{r}F\,\partial _{r}G\right] -e^{-2F}
\end{equation}

\bigskip

\begin{equation}
G^{2}\,_{2}=e^{-2\Phi }\left[ \partial _{r}\partial _{r}\left( \Lambda
+G\right) -\partial _{r}\Phi \,\partial _{r}\left( \Lambda +G\right) +\left(
\partial _{r}\Lambda \right) ^{2}+\left( \partial _{r}G\right) ^{2}+\partial
_{r}\Lambda \,\partial _{r}G\right]
\end{equation}

\bigskip

\begin{equation}
G^{3}\,_{3}=e^{-2\Phi }\left[ \partial _{r}\partial _{r}\left( F+\Lambda
\right) -\partial _{r}\Phi \,\partial _{r}\left( F+\Lambda \right) +\left(
\partial _{r}\Lambda \right) ^{2}+\left( \partial _{r}F\right) ^{2}+\partial
_{r}F\,\partial _{r}\Lambda \right]
\end{equation}

\bigskip

\begin{equation}
G^{1}\,_{3}=G^{2}\,_{3}=G^{0}\,_{3}=G^{0}\,_{2}=G^{0}\,_{1}=0
\end{equation}

\section{Energy-momentum tensor of the Born-Infeld monopole}

In the tetrad defined by (10), the energy-momentum tensor takes a diagonal
form, being its components the following
\begin{equation}
-T_{00}=T_{11}=\frac{b^{2}}{4\pi }\left( \frac{\mathbb{R}-1}{\mathbb{R}}%
\right)
\end{equation}
\begin{equation}
T_{22}=T_{33}=\frac{b^{2}}{4\pi }\left( 1-\mathbb{R}\right)
\end{equation}
where we defined
\begin{equation}
\mathbb{R}\equiv \sqrt{1-\left( \frac{F_{01}}{b}\right) ^{2}}
\end{equation}
in this manner, one can see from the Einstein equation (18) the
characteristic property of the spherically symmetric space-times
\begin{equation}
G^{1}\,_{2}=-e^{-\left( F+G\right) }\frac{\cos \theta }{\sin \theta }%
\partial _{r}\left( G-F\right) =0\,\ \ \ \ \Rightarrow \,\ \ \ G=F
\end{equation}

\section{Equations for the electromagnetic fields of Born-Infeld in the
tetrad}

The equations that describe the dynamic of the electromagnetic fields of
Born-Infeld in a curved spacetime are
\begin{equation}
\nabla _{a}\mathbb{F}^{ab}=\nabla _{a}\left[ \frac{F^{ab}}{\mathbb{R}}+\frac{%
P}{b^{2}\mathbb{R}}\widetilde{F}^{ab}\right] =0\,\ \ \ \ \ \ \ \ \ \ \ \ \ \
\ \ \ \ \ \ \ \ \left( field\,equations\,\right)
\end{equation}
\begin{equation}
\nabla _{a}\,\,\widetilde{F}^{ab}=0\,\ \ \ \ \ \ \ \ \ \ \ \ \ \ \ \ \ \ \ \
\ \ \ \ \ \ \ \ \ (\ Bianchi^{\prime }s\,\ identity)\ \ \ \ \ \ \ \ \ \ \ \
\ \ \ \ \ \ \ \ \ \ \ \ \ \ \ \ \ \ \ \ \ \
\end{equation}
where we was defined
\begin{equation}
P\equiv -\frac{1}{4}F_{\alpha \beta }\widetilde{F}^{\alpha \beta }
\end{equation}
\begin{equation}
S\equiv -\frac{1}{4}F_{\alpha \beta }F^{\alpha \beta }
\end{equation}
\begin{equation}
\mathbb{R}\equiv \sqrt{1-\frac{2S}{b^{2}}-\left( \frac{P}{b^{2}}\right) ^{2}}
\end{equation}
The above equations can be expressed geometrically of the following manner
\begin{equation}
dF=0
\end{equation}
\begin{equation}
d^{\ast }\mathbb{F}=0
\end{equation}
in explicit form
\begin{equation}
dF=d\left( F_{01}\omega ^{0}\wedge \omega ^{1}\right) =\partial _{\theta
}\left( e^{\Lambda +\Phi }F_{01}\right) \,d\theta \wedge dt\wedge
dr=0\Rightarrow F_{01}=A\left( r\right)
\end{equation}

\bigskip

\begin{eqnarray}
d^{\ast }\mathbb{F} &=&d\left( -\mathbb{F}_{01}\omega ^{3}\wedge \omega
^{2}\right) =\partial _{r}\left( -e^{F+G}\mathbb{F}_{01}\sin \theta \right)
\,dr\wedge d\varphi \wedge d\theta =0  \notag \\
&\Rightarrow &e^{2G}\mathbb{F}_{01}=f
\end{eqnarray}
where $f$ is a constant. We can see from equation (26) and (28) that
\begin{equation*}
\mathbb{F}_{01}=\frac{F_{01}}{\sqrt{1-\left( \overline{F}_{01}\right) ^{2}}}
\end{equation*}
where we obtain the following form for the electric field of the
self-gravitating B-I monopole
\begin{equation}
F_{01}=\frac{b}{\sqrt{\left( \frac{b}{f}e^{2G}\right) ^{2}+1}}
\end{equation}
we can to associate$^{1}$%
\begin{equation}
f=br_{0}^{2}\equiv Q\,\ \ \ \ \Rightarrow \,\ \ \ F_{01}=\frac{b}{\sqrt{%
\left( \frac{e^{G}}{r_{0}}\right) ^{4}+1}}
\end{equation}
Where $r_{0}$ is a constant with units of longitude that in reference$^{1}$
was associated to the radius of the electron.

\section{System of equations for the self-gravitating BI-monopole}

From the Einstein equations we have
\begin{equation}
F=G
\end{equation}
\begin{equation*}
G^{0}\,_{0}=e^{-2\Phi }\left[ 2\,\partial _{r}\partial _{r}G-\partial
_{r}\Phi \,2\,\partial _{r}G+3\left( \partial _{r}G\right) ^{2}\right]
-e^{-2G}
\end{equation*}
\begin{equation}
G^{1}\,_{1}=e^{-2\Phi }\left[ 2\,\partial _{r}\Lambda \,\partial
_{r}G+\,\left( \partial _{r}G\right) ^{2}\right] -e^{-2G}
\end{equation}
\begin{equation}
G^{2}\,_{2}=G^{3}\,_{3}=e^{-2\Phi }\ \Xi
\end{equation}
\begin{equation*}
\Xi \equiv \left[ \partial _{r}\partial _{r}\left( \Lambda +G\right)
-\partial _{r}\Phi \,\partial _{r}\left( \Lambda +G\right) +\left( \partial
_{r}\Lambda \right) ^{2}+\left( \partial _{r}G\right) ^{2}+\partial
_{r}\Lambda \,\partial _{r}G\right]
\end{equation*}
and the components of the energy-momentum tensor takes its explicit form
reemplacing the $F_{01}$ that we was found in equations (38)
\begin{equation}
-T_{00}=T_{11}=\frac{b^{2}}{4\pi }\left( 1-\sqrt{\left( \frac{r_{0}}{e^{G}}%
\right) ^{4}+1}\right)
\end{equation}
\begin{equation}
T_{22}=T_{33}=\frac{b^{2}}{4\pi }\left( 1-\frac{1}{\sqrt{\left( \frac{r_{0}}{%
e^{G}}\right) ^{4}+1}}\right)
\end{equation}
and from
\begin{equation*}
G^{a}\,_{b}=8\pi T^{a}\,_{b}
\end{equation*}
we obtains the components of the Einstein tensor for our problem.

\subsection{Reduction and solutions of the system of Einstein-Born-Infeld
equations}

Of the above expressions, we can see that $G^{0}\,_{0}=G^{1}\,_{1},$then
\begin{equation}
\partial _{r}\partial _{r}G+\left( \partial _{r}G\right) ^{2}-\partial
_{r}G\partial _{r}\left( \Phi +\Lambda \right) =0
\end{equation}
This equation (44) \ we can to reduce it of following manner. First we make
\begin{equation}
\partial _{r}G\equiv \xi
\end{equation}
with this change of variables, in the equation (44) we have first
derivatives only
\begin{equation}
\partial _{r}\xi +\xi ^{2}-\xi \partial _{r}\left( \Phi +\Lambda \right) =0
\end{equation}
dividing the above expression (46) by $\xi $ and making the substitution
\begin{equation}
\chi \equiv \ln \xi
\end{equation}
we have been obtained the following inhomogeneous equation
\begin{equation}
\partial _{r}\chi +e^{\chi }=\partial _{r}\left( \Phi +\Lambda \right)
\end{equation}
the homogeneus part of the last equation is easy to integrate
\begin{equation}
\chi _{h}=-\ln r
\end{equation}
Now, of the usual manner, we makes in (48) the following substitution
\begin{equation}
\chi =\chi _{h}+\chi _{p}=-\ln r+\ln \mu =-\ln r+\ln \left( 1+\eta \right)
\end{equation}
then
\begin{eqnarray}
\partial _{r}\ln \left( 1+\eta \right) +\frac{\eta }{r} &=&\partial
_{r}\left( \Phi +\Lambda \right) \Rightarrow \\
\partial _{r}\left[ \ln \left( 1+\eta \right) +\mathcal{F}\left( r\right)
-\left( \Phi +\Lambda \right) \right] &=&0  \notag \\
\ln \left( 1+\eta \right) +\mathcal{F}\left( r\right) -\left( \Phi +\Lambda
\right) &=&cte=0  \notag
\end{eqnarray}
where$\ \ \frac{d\mathcal{F}\left( r\right) }{dr}\equiv \frac{\eta (r)}{r}$.
The constant must be put equal to zero for to obtains the correct limit.
Finally the form of the exponent $G$ is
\begin{equation}
G=\ln r+\mathcal{F}\left( r\right)
\end{equation}
The next step is to put $\Phi $ in function of $\Lambda $ and $G$ in the
expression (39). After of tedious but straighforward computations and
integrations, we obtain
\begin{equation}
e^{2\Lambda }=1+a_{0}\,e^{-G}+e^{2G}\frac{2b^{2}}{3}-2b^{2}e^{-G}\int^{Y%
\left( r\right) }\sqrt{Y^{4}+\left( r_{0}\right) ^{4}}dY
\end{equation}
Where, we defined
\begin{equation*}
Y(r)=e^{G}
\end{equation*}
and $a(0)$ is an integration constant.

Hitherto, we know that $\mathcal{F}$ is an arbitrary function of the radial
coordinate $r$\ , but for to be sure of it, we must to introduce the fuction
$\Lambda $\ given for above equation, in the Einstein equations (41) and to
verified that $G_{22}=G_{33}$. Successfully, this equality is verified and
the functions $\Lambda ,\Phi $ and $G$ remains matematically determinate. In
this manner the line element of our problem (7) takes the following form
\begin{equation}
ds^{2}=-e^{2\Lambda }dt^{2}+e^{2\mathcal{F}\left( r\right) }\left[
e^{-2\Lambda }\left( 1+r\,\ \partial _{r}\mathcal{F}\left( r\right)
\,\right) ^{2}dr^{2}+r^{2}\left( d\theta ^{2}+\sin ^{2}\theta \,d\varphi
^{2}\right) \right]
\end{equation}

\subsection{Analysis of the function $\mathcal{F}\left( r\right) $ from the
physical point of view$.$}

The function $\mathcal{F}\left( r\right) $ must to have the behaviour in the
form that the electric field of the configuration obey the following
requirements for gives a regular solution in the sense that was given by B.
Hoffmann and L. Infeld$^{3}$
\begin{equation}
\left. F_{01}\right| _{r=r_{o}}<b
\end{equation}
\begin{equation}
\left. F_{01}\right| _{r=0}=0
\end{equation}
\begin{equation}
\,\ \ \ \ \ \ \ \ \ \ \ \ \ \ \ \ \ \ \ \ \ \ \ \ \ \ \left. F_{01}\right|
_{r\rightarrow \infty }=0\,\ \ \ \ \ \ \ \ \ \ \ \ \text{assymptotically \
Coulomb}
\end{equation}
the simplest function $\mathcal{F}\left( r\right) $ that obey the above
conditions, is of the type
\begin{equation}
e^{2\mathcal{F}\left( r\right) }=\left[ 1-\left( \frac{r_{0}}{a\left|
r\right| }\right) ^{n}\right] ^{2m}
\end{equation}
where $a$ is an arbitrary constant, and the exponents $n$ and $m$ will obey
the following relation
\begin{equation}
mn>1\,\ \ \ \ \left( m,n\in \mathbb{N}\right)
\end{equation}
with
\begin{equation*}
0<a<1\ \text{or}\ -1<a<0\ \ \
\end{equation*}
depending on $m\left( n\right) $ is even or odd and
\begin{equation*}
a\neq 0
\end{equation*}
that put in sure a consistent regularization condition not only for the
electric (magnetic) field but for the energy-momentum tensor (42) and (43)
and the line element (54).

The analysis of the Riemann tensor indicate us that it is regular every
where and its components goes faster than $\frac{1}{r^{3}}$ when $%
r\rightarrow \infty $. With all this considerations, the metric solution to
the problem is
\begin{equation*}
ds^{2}=-e^{2\Lambda }dt^{2}+\left[ 1-\left( \frac{r_{0}}{a\left| r\right| }%
\right) ^{n}\right] ^{2m}\left\{ e^{-2\Lambda }dr^{2}\left[ \frac{1-\left(
\frac{r_{0}}{a\mid r\mid }\right) ^{n}\left( mn-1\right) }{\left[ 1-\left(
\frac{r_{0}}{a\mid r\mid }\right) ^{n}\right] }\right] ^{2}+\right.
\end{equation*}
\begin{equation}
\left. +r^{2}\left( d\theta ^{2}+\sin ^{2}\theta \,d\varphi ^{2}\right)
\right\}
\end{equation}
and the electric field takes the form
\begin{equation}
F_{01}=\frac{b}{\sqrt{1+\left[ 1-\left( \frac{r_{0}}{a\left| r\right| }%
\right) ^{n}\right] ^{4m}\left( \frac{r}{r_{0}}\right) ^{4}}}
\end{equation}
Note that if in the condition (55) we take $a=1$ and $\left. F_{01}\right|
_{r=r_{o}}=b$ (limiting value for the electric field in BI theory) the
energy momentum diverges automatically at $r=r_{0}$. Strictely, the
regularity conditions for the energy-momentum tensor (without divergences in
the ``border'' $r_{0}$ of the spherical configuration source of the
non-linear electromagnetic field) are
\begin{equation*}
\left. T_{ab}\right| _{r=r_{o}}=finite\ \ \ \ \ \ \Rightarrow \ \ \ \ \ \
-1<a<0\ \ or\ \ 0<a<1\ \ \ \
\end{equation*}
depending on parity of $m,$ $n$; and
\begin{equation*}
\left. T_{ab}\right| _{r=0}\rightarrow 0\ \ \ \ \ \ \ \Rightarrow \ \ \ \ \
\ \ \mathbb{R}\rightarrow 1\ \ \ \ \ \
\end{equation*}
For the magnetic monopole case the line element is as expression (60) with
the following obvious definition
\begin{equation*}
br_{0}^{2}\equiv Q_{m}
\end{equation*}
The magnetic field takes the following form
\begin{eqnarray*}
F_{23} &=&\frac{b}{\left[ 1-\left( \frac{r_{0}}{a\left| r\right| }\right)
^{n}\right] ^{2m}\left( \frac{r}{r_{0}}\right) ^{2}} \\
&=&\frac{Q_{m}}{\left[ 1-\left( \frac{r_{0}}{a\left| r\right| }\right) ^{n}%
\right] ^{2m}r^{2}}
\end{eqnarray*}
and the considerations about the regularity conditions on the energy
momentum tensor is as the electric monopole case.

\subsection{\protect\bigskip Interesting cases for particular values of $n$
and $m$}

Because
\begin{equation*}
\exp 2\mathcal{F}\left( r\right) =\left[ 1-\left( \frac{r_{0}}{a\left|
r\right| }\right) ^{n}\right] ^{2m}
\end{equation*}
is easy to see that for $m=0$%
\begin{equation*}
e^{G}=r
\end{equation*}
and we obtains the spherically symmetric line element of Hoffmann$^{2}$ and
the electric field $F_{01}$and the energy-momentum tensor $\ T_{ab\text{ }}$%
take the form of the well know EBI solution for the electromagnetic geon of
Demi\'{a}nski$^{4}$ .

By other hand, in the \textit{limit }when: $a\rightarrow 1$, $n\rightarrow 4$
and $m\rightarrow \frac{1}{4}$ we have
\begin{equation*}
F_{01}\rightarrow \frac{b}{\sqrt{1+\left[ 1-\left( \frac{r_{0}}{\left|
r\right| }\right) ^{4}\right] \left( \frac{r}{r_{0}}\right) ^{4}}}=\frac{Q}{%
r^{2}}
\end{equation*}
where (as is usually taked) $br_{0}^{2}\equiv Q$ . How we see, we obtain as
solution in the \textit{limit} the Maxwellian linear field. Note that the
values of $a$ and the\ exponents $m$ and $n$ are restricted by conditions
(59).

\subsection{Analysys of the metric}

We have the metric (54)
\begin{equation*}
ds^{2}=-e^{2\Lambda }dt^{2}+e^{2\mathcal{F}\left( r\right) }\left[
e^{-2\Lambda }\left( 1+r\,\ \partial _{r}\mathcal{F}\left( r\right)
\,\right) ^{2}dr^{2}+r^{2}\left( d\theta ^{2}+\sin ^{2}\theta \,d\varphi
^{2}\right) \right]
\end{equation*}
if we make the substitution
\begin{equation*}
Y\equiv r\,e^{\mathcal{F}\left( r\right) }
\end{equation*}
and differentiating it
\begin{equation*}
dY\equiv \,e^{\mathcal{F}\left( r\right) }\left( 1+r\,\ \partial _{r}%
\mathcal{F}\left( r\right) \,\right) dr
\end{equation*}
the interval (7) takes the form
\begin{equation*}
ds^{2}=-e^{2\Lambda }dt^{2}+e^{-2\Lambda }dY^{2}+Y^{2}\left( d\theta
^{2}+\sin ^{2}\theta \,d\varphi ^{2}\right)
\end{equation*}
we can see that the metric (in particular the $g_{tt}$ coefficient), in the
new coordinate $Y(r)$, takes the similar form like a Demianski solution for
the Born-Infeld monopole spacetime :
\begin{equation*}
e^{2\Lambda }=1-\frac{2M}{Y}-\frac{2b^{2}r_{o}^{4}}{3\left( \sqrt{%
Y^{4}+r_{o}^{4}}+Y^{2}\right) }-\frac{4}{3}b^{2}r_{o}^{2}\,_{2}F_{1}\left[
1/4,1/2,5/4;-\left( \frac{Y}{r_{0}}\right) ^{4}\right]
\end{equation*}
here $M$ is an integration constant, which can be interpreted as an
intrinsic mass, and $_{2}F_{1}$ is the Gauss hypergeometric function$^{14}$.
We have pass
\begin{equation*}
g_{rr}\rightarrow g_{YY},\,\ \ \ \ \ \ \ \ \ \ \ \ \ \ g_{tt}\left( r\right)
\rightarrow g_{tt}\left( Y\right)
\end{equation*}
Specifically, for the form of the $\mathcal{F}\left( r\right) $ given by
(58), $Y$ is
\begin{equation*}
Y^{2}\equiv \left[ 1-\left( \frac{r_{o}}{a\left| r\right| }\right) ^{n}%
\right] ^{2m}r^{2}
\end{equation*}
Now, with the metric coefficients fixed to a asymptotically Minkowskian
form, one can study the asymptotic behaviour of our solution. A regular,
asymptotically flat solution with the electric field and energy-momentum
tensor both regular, in the sense of B. Hoffmann and L. Infeld is when the
exponent numbers of $Y(r)$ take the following particular values:
\begin{equation*}
n=3\ \ \ and\ \ \ \ \ m=1
\end{equation*}
In this case, and for $r>>r_{0}$, we have the following assymptotic
behaviour
\begin{equation*}
e^{2\Lambda }=1-\frac{2M}{r}-\frac{8b^{2}r_{o}^{4}K\left( 1/2\right) }{%
3r_{o}r}-2\frac{b^{2}r_{o}^{4}}{r^{2}}+...
\end{equation*}
A distant observer will associate with this solution a total mass
\begin{equation*}
M_{eff}=M+\frac{4b^{2}r_{o}^{4}K\left( 1/2\right) }{3r_{o}}
\end{equation*}
and total charge
\begin{equation*}
Q^{2}=2b^{2}r_{o}^{2}
\end{equation*}
Note that when the intrinsec mass $M$ is zero the line element is regular
everywhere, the Riemann tensor is also regular everywhere and hence the
space-time is singularity free. The electromagnetic mass
\begin{equation*}
M_{el}=\frac{4b^{2}r_{o}^{4}K\left( 1/2\right) }{3r_{o}}
\end{equation*}
and the charge $Q$ are the \textit{twice} that the electromagnetic charge
and mass of the Demianski solution$^{4}$ for the static electromagnetic
geon. Note that the $M_{el}$ is necessarily positive, which was not the case
in the Schwarzschild line element.

Is not very difficult to check that (for $m=1$ and $n=3$) the maximum of the
electric field (see figures) is not in $r=0$ , but in the \textit{physical}
\textit{border} of the spherical configuration source of the electromagnetic
fields. It means that $Y\left( r\right) $ maps correctely the internal
structure of the source in the same form that the quasiglobal coordinate of
the reference$^{15}$ for the global monopole in general relativity. The lack
of the conical singularities at the origin is because the very well
description of the manifold in the neighbourhood of $r=0$ given by $Y\left(
r\right) \,$. The function $Y\left( r\right) $ for the values of the $m$ and
$n$ parameters given above, have two differents behaviours near of $\ r=0$
depending of the $a$ sign \
\begin{equation*}
\text{for\ }a<0\text{ \ \ \ \ \ \ \ \ when \ }r\rightarrow \infty ,\ \
Y\left( r\right) \rightarrow \infty \ \ \ \ \ \ \text{and\ when \ }%
r\rightarrow 0,\ Y\left( r\right) \rightarrow \infty
\end{equation*}
\begin{equation*}
\text{for\ }a>0\text{ \ \ \ \ \ \ \ \ when \ }r\rightarrow \infty ,\ \
Y\left( r\right) \rightarrow \infty \ \ \ \ \ \ \text{and\ when \ }%
r\rightarrow 0,\ Y\left( r\right) \rightarrow -\infty
\end{equation*}
Note that the case \ $a>0$ will be excluded because in any value $%
r_{0}\rightarrow $ $Y\left( r_{0}\right) =0$ , the electric field takes the
limit value $b$ and the condition (54) is violated. The metric (see figures)
and the energy-momentum tensor remains \textit{both }regulars at the origin
(it is: $g_{tt}\rightarrow -1,T_{\mu \nu }\rightarrow 0$ $\ $\ for $%
r\rightarrow 0$).

Because the metric is regular ($g_{tt}=-1,$ at $r=0$ and at $r=\infty $),
its derivative must change sign. In the usual gravitational theory of
general relativity the derivative of $g_{tt}$ is proportional to the
gravitational force which would act on a test particle in the Newtonian
approximation. In Einstein-Born-Infeld theory with this new static solution,
it is interesting to note that although this force is attractive for
distances of the order $r_{0}<<r$ , it is actually a repulsion for very
small $r$. For $r$ greater than $r_{0},$ the line element closely
approximates to the Schwarzschild form. Thus the regularity condition shows
that the electromagnetic and gravitational mass are the same and, as in the
Newtonian theory, we now have the result that the attraction is zero in the
center of the spherical configuration source of the electromagnetic field.

\section{Conclusions.}

In this report a \textit{new }exact solution of the Einstein-Born-Infeld
equations for a static spherically symmetric monopole is presented. The
general behaviour of the geometry, is strongly modified according to the
value that takes $r_{0}\ $(Born-Infeld radius$^{1,9}$) and three new
parameters: $a$, $m$ and $n$.

The fundamental feature of this solution is the lack of conical
singularities at the origin when: $-1<a<0$ \ or $0<a<1$ (depends on parity
of $m$ and $n$) and $mn>1$. In particular, for $m=1$ and $n=3$, with the
parameter $a$ in the range given above and the intrinsic mass of the system $%
M$ is zero, the strong regularity conditions given by B. Hoffmann and L.
Infeld in reference$^{3}$, holds in all the spacetime. For the set of values
for the parameters given above, the solution is asymptotically flat, free of
singularities in the electric field, metric, energy-momentum tensor and
their derivatives (with derivative values zero for $r\rightarrow 0$); and
the electromagnetic mass (ADM)\ of the system is a twice that the
electromagnetic mass of other well known [2, 10, 14] solutions for the
Einstein-Born-Infeld monopole. The electromagnetic mass $M_{el}$
asymptotically is necessarily positive, which was not the case in the
Schwarzschild line element.

This solution have a surprising similitude with the metric for the global
monopole in general relativity given in reference$^{15-16}$ in the sense
that the physic of the problem have a correct description only by means of a
new radial function $Y\left( r\right) $.

Because the metric is regular ($g_{tt}=-1,$ at $r=0$ and at $r=\infty $),
its derivative (that is proportional to the the force in Newtonian
approximation) must change sign. In Einstein-Born-Infeld theory with this
new static solution, it is interesting to note that although this force is
attractive for distances of the order $r_{0}<<r$ , it is actually a
repulsive for very small $r$.

With this new regular solution, we also show that more strong conditions are
needed for to solve the problem of the lack of uniqueness of the function
action in non-linear electrodynamics.

\section{Acknowledgements:}

I am very thankful to Professors Yu. Stepanovsky, G. Afanasiev and B.
Barbashov for their guide in my scientific formation and very useful
discussions. I am very grateful to all people of the Bogoliubov Laboratory
of Theoretical Physics and Directorate of JINR\ for their hospitality and
support.

\bigskip

\subsection{Appendix:{\protect\LARGE \ }connections and curvature forms from
the geometrical Cartan's formulation}

$\bigskip \cdot $Making the exterior derivatives of $\omega ^{\alpha }$ we
computing the connection 1-forms $\omega ^{\alpha }\,_{\beta }$:
\begin{equation*}
\omega ^{0}\,_{1}=\omega ^{1}\,_{0}=e^{-\Phi }\,\ \partial _{r}\Lambda \,\
\omega ^{0}\,
\end{equation*}
\begin{equation*}
\omega ^{2}\,_{1}=-\omega ^{1}\,_{2}=e^{-\Phi }\,\ \partial _{r}F\left(
r\right) \,\ \omega ^{2}\,
\end{equation*}
\begin{equation*}
\omega ^{3}\,_{1}=-\omega ^{1}\,_{3}=e^{-\Phi }\,\ \partial _{r}G\left(
r\right) \,\ \omega ^{3}\,
\end{equation*}
\begin{equation}
\omega ^{3}\,_{2}=-\omega ^{2}\,_{3}=\frac{\cos \theta }{sen\theta }%
e^{-F\left( r\right) }\omega ^{3}\,
\end{equation}
$\cdot $Making the exterior derivatives of $\omega ^{\alpha }\,_{\beta }$ we
computing the curvature 2-forms $\mathcal{R}^{\alpha }\,_{\beta }$:
\begin{equation*}
\mathcal{R}^{0}\,_{1}=e^{-2\Phi }\left( \partial _{r}\partial _{r}\Lambda
-\partial _{r}\Phi \,\partial _{r}\Lambda +\left( \partial _{r}\Lambda
\right) ^{2}\right) \,\omega ^{1}\wedge \omega ^{0}
\end{equation*}
\begin{equation*}
\mathcal{R}^{2}\,_{1}=e^{-2\Phi }\left( \partial _{r}\partial _{r}F-\partial
_{r}\Phi \,\partial _{r}F+\left( \partial _{r}F\right) ^{2}\right) \,\omega
^{1}\wedge \omega ^{2}
\end{equation*}
\begin{equation*}
\mathcal{R}^{3}\,_{2}=e^{-\left( F+\Phi \right) }\,\partial _{r}\left(
G-F\right) \,\frac{\cos \theta }{sen\theta }\,\omega ^{1}\wedge \omega
^{3}+\left( e^{-2\Phi }\partial _{r}G\,\,\partial _{r}F-e^{-2F}\right)
\,\omega ^{2}\wedge \omega ^{3}
\end{equation*}
\begin{eqnarray*}
\mathcal{R}^{3}\,_{1} &=&e^{-2\Phi }\left( \partial _{r}\partial
_{r}G-\partial _{r}\Phi \,\partial _{r}G+\left( \partial _{r}G\right)
^{2}\right) \,\omega ^{1}\wedge \omega ^{3}+ \\
&&+e^{-\left( F+\Phi \right) }\,\partial _{r}\left( G-F\right) \,\frac{\cos
\theta }{sen\theta }\,\omega ^{2}\wedge \omega ^{3}
\end{eqnarray*}
\begin{equation*}
\mathcal{R}^{0}\,_{2}=-e^{-2\Phi }\,\partial _{r}\Lambda \,\,\partial
_{r}F\,\,\omega ^{0}\wedge \omega ^{2}
\end{equation*}
\begin{equation}
\mathcal{R}^{0}\,_{3}=-e^{-2\Phi }\,\partial _{r}\Lambda \,\,\partial
_{r}G\,\,\omega ^{0}\wedge \omega ^{3}
\end{equation}

\begin{center}
\bigskip
\end{center}

\subsection{References}

\bigskip

\

$^{1}$ M. Born and L. Infeld, Proc. Roy. Soc.(London) \textbf{144}, 425
(1934).

$^{2}$ B. Hoffmann, Phys. Rev. \textbf{47}, 887 (1935).

$^{3}$ B. Hoffmann and L. Infeld, Phys. Rev. \textbf{51}, 765 (1937).

$^{4}$ M. Demianski, Found. Phys. Vol. 16 , No. 2, 187 (1986).

$^{5}$ D. Harari and C. Lousto, Phys. Rev. D \textbf{42}, 2626 (1990).

$^{6}$ M. Barriola and A. Vilenkin, Phys. Rev. Lett. \textbf{63}, 341 (1989).

$^{7}$ L. D. Landau and E. M. Lifshitz, \textit{Teoria Clasica de los Campos}%
, (Reverte, Buenos Aires, 1974), p. 574.

$^{8}$ C. Misner, K. Thorne and J. A. Wheeler, \textit{Gravitation},
(Freeman, San Francisco, 1973), p. 474.

$^{9}$ M. Born , Proc. Roy. Soc. (London) \textbf{143}, 411 (1934).

$^{10}$ R. Metsaev and A. Tseytlin, Nucl. Phys.\textbf{B 293}, 385 (1987).

$^{11}$ Yu. Stepanovsky, Electro-Magnitie Iavlenia \textbf{3}, Tom 1, 427
(1988).

$^{12}$ D. J. Cirilo Lombardo, Master Thesis, Universidad de Buenos Aires,
Argentina, 2001.

$^{13}$ S. Chandrasekhar, \textit{The Mathematical Theory of Black Holes},
(Oxford University Press, New York, 1992).

$^{14}$ A. S. Prudnikov, Yu. Brychov and O. Marichev, \textit{Integrals and
Series }(Gordon and Breach, New York, 1986).

$^{15}$ K. A. Bronnikov, B. E. Meierovich and E. R. Podolyak, JETP\ \textbf{%
95}, 392 (2002).

$^{16}$ D. J. Cirilo Lombardo, Preprint JINR-E2-2003-221.

\bigskip

\bigskip

\bigskip

\bigskip

\bigskip

\bigskip

\end{document}